# Implications for $^{14}$C Dating of the Jenkins-Fischbach Effect and Possible Fluctuation of the Solar Fusion Rate

Alvin J. Sanders
University of Tennessee, Knoxville, TN, USA
**asanders@utk.edu**
Mailing address: Oak Ridge National Laboratory, MS 6054, Oak Ridge, TN 37831

It has long been known that the $^{14}$C calibration curve, which relates the known age of tree rings to their apparent $^{14}$C ages, includes a number of "wiggles" which clearly are not experimental errors or other random effects. A reasonable interpretation of these wiggles is that they indicate that the Sun's fusion "furnace" is pulsating, perhaps for reasons similar to that of the Cepheid variables, albeit under a very different regime of pressure and temperature. If this speculation is correct, we are seeing the heartbeat of the Sun—the $^{14}$C calibration curve is the Sun's "neutrino-cardiogram."

Elevated neutrino flux during a relatively brief period would have two effects: (1) a surge in $^{14}$C fraction in the atmosphere, which would make biological samples that were alive during the surge appear to be "too young" (2) depletion of $^{14}$C in the biotic matter already dead at the time of the surge; this is a consequence of the recently discovered Jenkins-Fischbach effect, which is an observed correlation between nuclear decay rates and solar activity or Earth-Sun distance.

In addition, the precise value at any given time of the "half-life" of any unstable isotope—including $^{14}$C—must now be considered in doubt, since the Jenkins-Fischbach effect implies that we may no longer view the decay rate of an isotope as intrinsically governed and therefore a constant of Nature.

Introduction

In 1960 Willard F. Libby received the Nobel Prize for the breakthrough idea of carbon-14 dating. Libby realized that all living matter in the biosphere would be in approximate equilibrium with the atmosphere with respect to the proportions of $^{14}$C and $^{12}$C, while in all dead biota the $^{14}$C fraction would decrease according to the known law of radioactive decay, $N(t) = N_o \exp(-t \, Ln \, 2/T)$, where T is the half life. Thus, the "$^{14}$C age" of any dead plant or animal matter was obtained simply by solving for the decay time which was required for the $^{14}$C fraction in the dead biological sample to decline from the (presumably constant) atmospheric value to the experimental value of the sample:

$$t = -(T/Ln \, 2) \, Ln[N(t)/N_o] \quad (1)$$

Systematic deviations from this simple law were discovered in the 1960s and early 1970s through dendrochronological studies of very long-lived trees such as Giant Sequoias and Bristlecone Pines. Since the ages of their tree rings were obviously known precisely (both for living trees and for dead samples for which dendrochronological matches could be found), comparing these ages against their $^{14}$C ages yielded a calibration curve for adjusting the simple $^{14}$C ages (Eq. 1). For samples more than 3000 years old, the $^{14}$C ages substantially understate the actual ages as determined by dendrochronology. The resulting corrections yielded a vast improvement in dating accuracy and thus opened an entire new chapter in radiocarbon dating (Ralph & Michael, 1974).

The non-equality of $^{14}$C and dendrochronological ages has been implicitly assumed to be due solely to the fact that the atmospheric fraction of $^{14}$C has not been constant—the default assumption underlying Eq. 1—but rather has varied in time. In turn, the variation in atmospheric $^{14}$C has been widely attributed chiefly to variations in the magnetic field of the Earth (for a review, see Chiu et al., 2007). This interpretation of course takes no account of the Jenkins-Fischbach effect (2008a & 2008b), which introduces another mechanism for the discrepancy between dendrochronological and $^{14}$C ages. Although considerable success has been achieved in explaining the large-scale effects of the $^{14}$C calibration curve, effects that are of small scale in time and/or magnitude, as described in the following section, have not been satisfactorily explained (Chiu et al., 2007).

Wiggles in $^{14}$C calibration curve

Unfortunately, the $^{14}$C calibration curve suffers from a major complication: It has a number of "wiggles." With respect to the objective of correcting the $^{14}$C dates, at best these wiggles are annoying and—far worse—the wiggles mean that the calibration curve is *not single-valued* when the objective is to obtain the dendrochronological date that corresponds to a given $^{14}$C date. Discovery of the wiggles has been credited to Hessel de Vries (Taylor, 2000) and to Hans Seuss (see for example, Suess & Linick, 1990), whose La Jolla laboratory was the leader in precision in the early days of radiocarbon dating. The term "Seuss wiggles" seems to have gained widespread acceptance.

For a number of years there was debate within the radiocarbon community about whether the Seuss wiggles were real or merely statistical fluctuations. Various investigators concluded that they were merely statistical artifacts (Ottaway & Ottaway, 1974; Clark, 1975). It is now clear in hindsight that these wiggles are not experimental errors or other random effects (Suess & Linick, 1990). Moreover, it is now generally accepted that the fluctuation has a pronounced Fourier component at a period of ~200 years (Kruse/Suess, 1980). Thus, there is now a broad consensus that these wiggles are indeed real.

A further compelling reason for believing that the wiggles are real (not yet published to the author's knowledge) is that they tend to have a *characteristic shape* (Please consult Figure 2 of Ralph & Michael, 1974). To wit:

- The negative excursions ($^{14}$C ages younger than trend line) are sharply pointed.
- The positive excursions ($^{14}$C ages older than trend line) are broad and fairly smooth.

This pairing is not consistent with stochastic processes but is consistent with some kind of process that is governed by something similar to a relaxation oscillator, which will cause periodic surges in the number of high-energy neutrons to feed the reaction

$$^{14}N + n \rightarrow {}^{14}C + {}^{1}H \qquad (2)$$

in the atmosphere. It seems unlikely that neutrons in sufficient quantity for such a surge could be produced in the atmosphere by the Solar wind, since it consists mostly of *charged* particles, so any surge would be accompanied by a spectacular auroral display (and any Solar-wind neutrons originating at the Sun would also decay to protons before reaching the Earth). Such displays would not have gone unnoticed, as evidenced by the giant storm of 1859 (Oldenwald & Green, 2008). In particular, Robert H. Dicke or Robert R. Newton would likely have recorded any such events during their searches of Medieval records for evidence of the Nordtvedt effect. If so, they did not emphasize it in their subsequent writings (Dicke, 1966; Newton, 1970 and 1972). In short, the absence of evidence of giant auroral displays at intervals of ~200 yr suggests that the surges in $^{14}$C production are initiated by some *neutral* particle. Until now most physicists would not consider neutrinos as plausible candidates, since they interact only weakly. However, the newly-discovered Jenkins-Fischbach effect may force reconsideration of the role of neutrinos in nuclear decays and, possibly, other nuclear processes. In particular, neutrinos in large numbers may now be attractive candidates for either initiating the reaction of Eq. 2 or otherwise producing $^{14}$C.

> The author therefore suggests that the wiggles in the $^{14}$C calibration curve may be due chiefly to variations in Solar neutrino output, as implied by the work of Jenkins and Fischbach; that the rate of nuclear fusion in the Sun's core may be pulsing with a quasi-period of ~200 years, which would account for most of the major wiggles in the $^{14}$C calibration curve; and that neutrino bursts from large solar storms are candidates as explanations of some of the minor wiggles in the calibration curve.

Such a pulsation might possibly be due to the same kind of processes which account for the fluctuation of the rate of fusion in Cepheid variable stars, albeit in a completely different regime of temperature and pressure. The periods of Cepheids are typically a matter of days to weeks, rather than ~200 years for the $^{14}$C wiggles. The light output of a Cepheid fluctuates because its hydrogen is so nearly exhausted that most of the fusion is taking place fairly close to the stellar surface, and the thin outer layer of the star is simply not heavy enough to "keep the lid on" and permit anything approaching uniformity. Thus, their "fusion furnaces" are not stable but rather undergo a cycle of alternating phases of gradual cooling/compression followed by rapid heating/expansion.

Conversely, this model argues that the reason an ordinary star is *not* variable is that its fusion furnace is so deep in the interior that most of the mass of the star is available to press down on the fusion furnace and, hence, keep the fusion rate essentially uniform. A caveat about this picture of an ordinary star is that, since we cannot observe its interior directly, if cyclical behavior were in fact typical of ordinary stars too, we would have no way of knowing it until now. Since the photons from the fusion furnace are quickly thermalized and then require ~$10^6$ years to complete their random walk to the stellar surface, significant variability in the fusion rate could be consistent with nearly uniform photon output at the stellar surface. On the other hand, neutrinos—unlike photons—escape immediately, thus providing a real-time signature of the activity of the fusion furnace at the core, if indeed they have observable effects in the $^{14}$C calibration curve. The cycle of sharp minima and broad maxima in the $^{14}$C calibration curve of Ralph & Michael suggests that a quasi-cyclical pattern in the fusion rate within the Sun's core does indeed exist, and that a full cycle consists of a relatively long interval with gradually decreasing fusion rate followed by a sharp—not to say convulsive—jump in the rate of fusion which persists for a fairly brief period.

Moreover, it seems unlikely that *anything other than the Sun* could be the source of sufficient neutrino flux to influence the $^{14}$C calibration curves, chiefly because other sources are too distant and/or lack the observed periodicity (see Appendix).

Subsequent (post-1974) $^{14}$C Calibration Curves

After the original $^{14}$C calibration curve of Ralph and Michael, a number of subsequent calibration curves have been published (Klein, et al., 1982; Klein, et al., 1982; Stolk, et al., 1994; Stuiver, 1993; Pearson & Stuiver, 1993; Pearson & Qua, 1993; Stuiver & Pearson, 1993; Stolk, et al., 1994; Stuiver et al., 1998; Reimer et al., 2004), most importantly those which represented the consensus findings of the various leading laboratories. It is striking that, in these later calibration curves, the wiggles often lack the distinctive sharp-tipped minima that were so prominent in the 1974 curve.

The figure below represents an attempt by the author to model the wiggles—with their sharply-tipped minima—assuming that the cause could be a supernova, using Kepler's supernova as an example (As in Ralph & Michael (1974), 9-point smoothing is used.). Although supernovae are in fact too distant to account for the wiggles, the graph nevertheless gives a useful qualitative picture of the shape of the wiggles as shown in Figure 2 of Ralph & Michael (1974). This distinctive shape was one of the most compelling reasons to believe that the wiggles are due to some characteristic physical process.

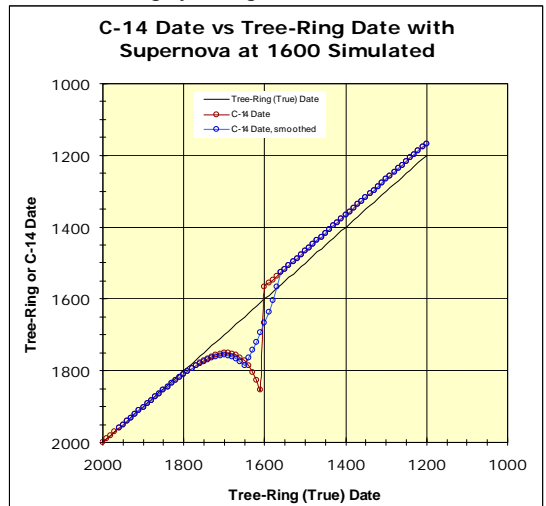

A possible reason why the sharp minima of the wiggles disappeared in later calibration curves is that the $^{14}$C community has generally viewed the wiggles as chiefly a nuisance, in that the wiggles obstructed their overarching goal of extracting precise true dates from $^{14}$C ages. The issue of how smooth the wiggles ought to be is explicitly discussed in a number of recent papers (Suess & Linick, 1990; Törnqvist & Bierkens, 1994). Thus, it is possible that, in their quest to suppress the wiggles as much as possible, various authors devised strongly-smoothing algorithms after 1974, which had the unintended effect of suppressing the asymmetry in the wiggles.

At least three kinds of evidence support this interpretation for the unexpected smoothness and symmetry of wiggles in post-1974 calibrations:
 (a) The process known as "windsorizing" data was used, at least in 1980, to suppress outliers. In this process, for any $^{14}$C age that was more than a 2-SD (standard deviation) outlier, the resulting data point was removed and replaced by a synthetic point that was exactly a 2-SD outlier (Klein, Lerman, Damon and Linick, 1980).
(b) Even H.E. Suess, who for years was the leader in high-precision $^{14}$C analysis and the only investigator to see the wiggles, has explicitly stated his philosophical conviction—in 1990—that the wiggles simply had to be intrinsically smooth: *natura non facet saltum* (Nature does not make jumps) (Suess & Linick, 1990).
(c) Moreover, the unsmoothed negative peaks might be distressingly large. For example, a really convulsive recovery of the fusion furnace of the Sun might result in a decade of $^{14}$C dates that are below the trend line (excessively young) by a matter of centuries. Any competent researcher, working with samples of known age that are less than 2000 years old, might view such $^{14}$C dates as simply "wrong". In this circumstance the researcher could—perhaps even *should*—reasonably attribute the "anomalous" result to contamination and discard the data without even recording the fact. This would of course result in a serious bias against the very data points that create the sharply-pointed minima. Thus, although there is valid scientific rationale for discarding such extreme "outliers," the unfortunate result is that a lot of valid data may have been discarded.

For completeness, we note that some $^{14}$C researchers have found positive value in the wiggles. A procedure known as "wiggle matching" has been used to achieve very precise dating of wood and charcoal samples when the approximate dates are known. This procedure entails matching the wiggles in the experimental sample to those in the calibration curve, very much like matching a key with its lock (see for example, Betancourt et al., 1999). More broadly, some researchers have emphasized the intrinsic value of the wiggles as windows into the history of carbon processes in the biosphere (see for example, Levin & Hesshaimer, 2000).

Further tests and clues to be sought

Any fluctuation of the Sun's fusion furnace might cause some variability in the area and/or temperature of the Sun's surface. Such changes could not be very large, or else a marked 200-year periodicity in climate would have already been observed (unless a concomitant change in the Sun's surface area would just offset its surface-temperature change). The threshold for causing noticeable climate changes is a fractional radius change of $\sim 10^{-3}$, since from the Stefan-Boltzmann law the resulting change in the mean temperature of the Earth would be $\sim$1K, other things being equal.

Past genetic changes may also hold clues in the form of episodes of elevated mutation rates. The state-of-the-art of paleo-genetic measurement may conceivably be sufficiently advanced now to provide time resolution of the rate of genetic change good enough to see a stair-step pattern with only a 200-year cycle (Here we refer to mutations *per se*, not to genetic drift and not to the natural selection process, although natural selection may of course follow from either mutations are genetic drift.).

Conclusion

The "wiggles" in the $^{14}$C calibration curve may be evidence of the variability of Solar neutrino flux at the Earth. Very importantly, the calibration curve suggests that the fusion rate in the Sun's core could be pulsing with a quasi-period of about 200 years. Such a pulsation would of course cause a quasi-periodic variation in neutrino flux. A surge in neutrino flux would have two effects:
- It would cause excess decays of the $^{14}$C isotopes in all dead biota (via the Jenkins-Fischbach effect), thus *increasing* their

apparent ages as indicated by their $^{14}$C ages."
• It would produce excess atmospheric $^{14}$C for a brief period, thus causing the biotic matter formed during the surge to look anomalously *young*—perhaps by very large amounts (which may have led to unwarranted discarding of good data).

These two effects together mean that the $^{14}$C calibration curve may constitute a "neutrino-cardiogram" giving a time history of the "heartbeat of the Sun."

**Appendix**
**Potential Non-Solar Sources of Neutrinos**

Heuristic considerations, including back-of-the-envelope calculations, show that it is implausible that anything other than the Sun could be the source of enough neutrinos to compete with the influence of Solar neutrinos.

<u>Other stars within the Milky Way as sources of neutrinos.</u>
The calculations below show that the fluxes due to the central bulge, the disk, and the local neighborhood of the Earth would all be approximately the same size if all stars had approximately the same neutrino output, and that the magnitude would be far below the Sun flux, in the absence of rather exotic processes in one or more of the galactic sources.

Let the Solar neutrino flux at the Earth be

Solar: $F_S = f_S/a^2$ (A1)

where $a$ is the distance to the Sun and $f_S$ is the neutrino output factor of the Sun.

The galactic neutrino flux at the Earth may be divided into three parts and estimated as follows:

Central bulge: $F_C \approx f_C \, n_C/R^2$ (A2)

where $f_C$ is the neutrino output factor per Solar mass of the central bulge, $n_C$ (~ $10^{11}$) is the mass of the central bulge in units of Solar mass (or, alternatively, the effective number of sun-like stars in the bulge), and $R$ (≈ 30,000 ly) is the distance of the Earth from the center of the galaxy. This result would be 7 or 8 orders of magnitude smaller than the Solar flux $F_S$ if the output factors were similar ($f_C \approx f_S$).

Disk: $F_D = \int f_D /r^2 \, (dn/dA) \, dA$ (A3)

where $f_D$ is the neutrino output factor per Solar mass of the disk, $dA$ is an area element in the disk, $r$ is the distance of the Earth from $dA$, and $(dn/dA)$ is the effective number of stars per unit area at $dA$ (all assumed to have roughly the mass of the Sun).

Local neighborhood: $F_L = \int f_L /r^2 \, (dn/dV) \, dV$ (A4)

where $f_L$ is the average neutrino output factor of the stars in a spherical volume surrounding the Earth, $dV$ is a volume element in this sphere, and $(dn/dV)$ is number of stars per unit volume at $dV$ (all assumed to have roughly the mass of the Sun). The local-neighborhood flux would of course be omni-directional (although not isotropic).

The local-neighborhood integral in Eq. A4 can be roughly approximated by pulling things out of the integral as follows:

$$\int f_L /r^2 \, (dn/dV) \, dV \sim <f_L \, dn/dV>  {}_0\!\int^b 1/r^2 \, 4\pi r^2 \, dr = [f_L \, n_D/(8\pi R^2 b)] \, 4\pi b = f_L n_D/(2R^2). \quad (A5)$$

Here $<f_L \, dn/dV>$ denotes an appropriate weighted average of the indicated quantities, $n_D$ is the total number of stars in the disk (~$10^{11}$ also), the local neighborhood is taken as a sphere around the Earth of radius b (b = approximately half the average thickness of the disk), the stellar density dn/dV in the disk is taken as $n_D$ divided by the total volume of the disk (taken as a cylinder of radius 2R and height 2b), and the stellar density within the local neighborhood is assumed to be average stellar density within the disk. The last assumption is plausible because the Earth is about half-way out from the center of the galaxy to the edge.

Finally, the disk integral in Eq. A3 can also be roughly approximated by pulling things out of the integral as follows:

$$\int f_D /r^2 \, (dn/dA) \, dA = <f_D \, dn/dA>  {}_b\!\int^R 1/r^2 \, 2\pi r \, dr = [f_D \, n_D/(4\pi R^2)] \, 2\pi \ln(R/b) \sim f_D \, n_D/R^2 \quad (A6)$$

Here $n_D$ (~$10^{11}$) is still the total number of stars in the disk, the areal stellar density dn/dA in the disk is taken as $n_D$ divided by the total area of the disk (radius 2R), but the integral is taken over an annulus *centered on the Earth* rather than the center of the galaxy and is therefore cut off at radius R. The inner radius of this annulus is taken as b; this is necessary both because the disk no longer looks very two-dimensional at distances less than b from the Earth and in order to avoid overlap between the integrals over the disk (Eq. A3) and over the local neighborhood (Eq. A4). The term ln(R/b) is approximately 2, since the radius of the galactic disk is 5 to 10 times its average thickness.

Recapping the flux due to the Sun and the three galactic regions:

| | | |
|---|---|---|
| Solar: | $F_S = f_S/a^2$, | (A1-redux) |
| Central bulge: | $F_C \approx f_C \, n_C/R^2$ | (A2-redux) |
| Disk: | $F_D \sim f_D \, n_D/R^2$ | (A6-redux) |
| Local neighborhood: | $F_L \sim f_L n_D/(2R^2)$ | (A5-redux) |

Thus, the three galactic components of galactic neutrino flux would be approximately equal if
(1) roughly half the stars were in the central bulge (i.e., $n_D \approx n_C$) and
(2) the average neutrino output per Solar mass were the about same in the three regions
(i.e., $f_L \approx f_D \approx f_C$).

The first assumption is sufficient for our purposes. The second may be wrong by multiple orders of magnitude. It would not be surprising if

$$f_L < f_D << f_C \quad (A7)$$

since the Earth is located in a very quiescent region of the galaxy and very violent processes are occurring at the core. In case $f_C$ were very extreme, to wit if

$$f_C / f_L \approx 10^8 \tag{A8}$$

for a few years, the surge could temporarily exceed the mean neutrino flux from the Sun.

One possible mechanism for a large $f_c$ might be a large black hole in a highly eccentric orbit around a super-massive black hole. The period could be ~200 years with appropriate parameters (E.g., semi-major axis ~75,000 AU (~1.2 ly) and mass ~$10^{10}$ Solar masses for the super-massive black hole). The mayhem caused by this pair when near closest approach might conceivably generate large numbers of neutrinos for a few years. However, any such effect would necessarily be produced by a tiny portion of the central bulge. The higher multipole moments of the orbital motion would be negligible at distances over a few light years. If Eq. A8 were somehow fulfilled on a steady-state basis it is even conceivable that *all nuclear decay processes could be predominantly stimulated* rather than governed by internal mechanisms within the nucleus. That is, there would be essentially no such thing as "natural" nuclear decay. Otherwise, Equations A-1, 2, 5, and 6 show that the neutrino flux from Sun dwarfs that from other stars within the Milky Way. In these arguments the author has assumed that considerations of neutrino flavor could not appreciably alter the result.

Supernovae as neutrino sources

The chief reason why the Sun dwarfs all other stars in terms of neutrino flux is that the Sun is so close to the Earth. This fact also would seem to rule out supernovae even more strongly. To wit, a naked-eye supernova is typically 10,000 ly ≈ $10^{20}$ km from the Earth, *vs.* 1 AU = $1.5 \times 10^8$ km, so the ratio of the radii squared suppresses a supernovae by a factor of ~$2 \times 10^{-24}$. Supernovae are of course vastly brighter— ~$10^{11}$ times the visible-photon brightness of a Sun-like star for a few days—which means that in 10 years (the typical averaging time for $^{14}$C data) a supernovae emits ~$10^8$ times as many photons as the Sun. Combining this with the distance effect gives ~ $(10^8) \times (2 \times 10^{-24})$ = ~$2 \times 10^{-16}$. That is, over 10 years the integrated flux of photons at the Earth from a typical naked-eye supernovae is about 16 orders of magnitude below the integrated photon flux from the Sun during the same period. Thus, only if the neutrino production were disproportionate to the photon flux by at least 14 orders of magnitude could a supernova sensibly affect the number of neutrinos received by the Earth.

Therefore, it seems implausible that supernovae could influence the $^{14}$C fraction in the atmosphere or in dead organic matter.

Other sources

The effect of distance suppression applies *a fortiori* to any extra-galactic source. E.g., Andromeda is ~$2 \times 10^6$ ly away, which is 3 orders of magnitude larger than the typical distances to naked-eye supernovae and to the center of our own galaxy.

Planets can also be ruled out because, although they are also very close to the Earth, we have no reason to suspect that they are periodically emitting bursts of neutrinos.

The Earth itself must not be ruled out prematurely. Conceivably some kind of quasi-periodic spasmodic Oklo reactor may exist deep within the Earth. It might be possible to rule this out by the non-observation of (1) any such effects during the previous controlled intra-terrestrial neutrino experiments, such as FermiLab's recent MINOS experiment, which entails beaming neutrinos to the Soudan iron mine in Minnesota, and (2) various magnetic or geological other effects that would result from any sort of giant swings in the Earth's core.

Additional considerations against non-Solar neutrino sources

• The 200-year regularity alone would also seem to rule out supernovae, which ought to be *randomly* distributed in time. Likewise stellar phenomena (even if one could somehow postulate a mechanism for correlated, laser-like, neutrino outbursts from groups of stars).
• A 200-year period is, to the author's knowledge, not compatible with radical changes in the Earth's magnetic field or the Sun's magnetic field. Moreover, navigators should have noticed any such episodes, especially within the past 550 years.
• All mechanisms based on extreme electromagnetic effects or enhanced Solar wind seem implausible, since any effect of this nature would cause dramatic auroras. The great magnetic storm of 1859 (Oldenwald & Green, 2008) shows that such effects can indeed be very dramatic. Nevertheless the 1859 storm cannot serve as a counter-example since, to the author's knowledge, other such storms have not been observed, and certainly not with a periodicity of ~200 years.

Conclusion of Appendix
It seems that only the Sun can plausibly be the source of whatever process is causing the wiggles in the $^{14}$C calibration curve.